\documentclass[aps,prl,twocolumn,showpacs,superscriptaddress]{revtex4-1}

\usepackage{graphicx}
\usepackage{dcolumn}
\usepackage{bm}
\usepackage{epstopdf}

\begin{document}

\newcommand{\bra}[1]{\left\langle#1\right|}
\newcommand{\ket}[1]{\left|#1\right\rangle}
\newcommand{\bracket}[2]{\big\langle#1 \bigm| #2\big\rangle}

\title{
  Kondo effect and spin quenching in high-spin molecules on metal substrates
}

\author{D. Jacob}
\email{djacob@mpi-halle.de}
\affiliation{Max-Planck-Institut f\"ur Mikrostrukturphysik, Weinberg 2, 06120 Halle, Germany}
\author{M. Soriano}
\affiliation{Departamento de F\'isica de la Materia Condensada, Universidad Aut\'onoma de Madrid, Madrid, Spain}
\author{J. J. Palacios}
\affiliation{Departamento de F\'isica de la Materia Condensada, Instituto de F\'isica de la Materia Condensada (IFIMAC), and Instituto Nicol\'as Cabrera (INC), Universidad Aut\'onoma de Madrid, Madrid, Spain}

\date{\today}

\pacs{75.50.Xx,72.10.Fk,71.27.+a,74.55.+v}

\begin{abstract}
  Using a state-of-the art combination of density functional theory and impurity solver techniques 
  we present a complete and parameter-free picture of the Kondo effect in the high-spin ($S=3/2$) 
  coordination complex known as Manganese Phthalocyanine adsorbed on the Pb(111) surface. We calculate 
  the correlated electronic structure and corresponding tunnel spectrum and find an asymmetric Kondo 
  resonance, as recently observed in experiments. Contrary to previous claims, the Kondo resonance 
  stems from only one of three possible Kondo channels with origin in the Mn $3d$-orbitals, 
  its peculiar asymmetric shape arising from the modulation of the hybridization due to a 
  strong coupling to the organic ligand. The spectral signature
  of the second Kondo channel is strongly suppressed as the screening occurs via the 
  formation of a many-body singlet with the organic part of the molecule. 
  Finally, a spin-1/2 in the $3d$-shell remains completely unscreened due to the lack of hybridization of the corresponding orbital 
  with the substrate, hence leading to a spin-3/2 underscreened Kondo effect.
\end{abstract}

\maketitle

Whenever a magnetic atom or magnetic molecule is coupled to metallic electrodes the conduction electrons
are likely to screen its magnetic moment through the Kondo effect \cite{kondo:ptp:64,hewson:book:97} which is
signaled by Fano-Kondo lineshapes in the conductance spectra 
\cite{madhavan:science:98,*li:prl:98,*knorr:prl:02,*yu:prl:05,*ternes:jphys:09}.
Although this usually occurs at low temperatures, it may have important consequences for possible 
applications of molecular magnets
\cite{gatteschi:book:06} as ultimately miniaturized magnetic storage units or as 
prospective nanoscale spintronics devices \cite{wolf:science:01}. 
Alternatively, the Kondo effect could also serve as a sensor 
of magnetic state changes when the molecule is subjected to mechanical deformation
\cite{iancu:nanolett:06,*choi:nanolett:10,parks:science:10} or chemical changes 
\cite{zhao:science:05,strozecka:prl:12,*Liu:SR:13} without the necessity of applying a magnetic field, thus 
opening the door to novel applications of this quantum effect. 

From a more fundamental point of view, atomic precision experimental control
offers the possibility to study a wide range of electron correlation phenomena 
related to the Kondo effect. For example, the atomic-scale control of atoms or 
molecules adsorbed on metal surfaces or anchored to nanoscopic electrodes allows 
for a direct manipulation of the orbital hybridization and for a controlled tuning 
from the so-called underscreened to the overscreened Kondo effects, both regimes 
showing interesting non-Fermi liquid behavior \cite{nozieres:jphys:80}.
Recently, underscreened Kondo effects have been reported for a C$_{60}$ quantum dot 
molecule coupled to metal leads \cite{roch:prl:09} and for a Co(tpy-SH)$_2$ complex 
coupled to Au nanocontacts \cite{parks:science:10}. The overscreened Kondo effect, 
on the other hand, has been very recently predicted to occur in Au nanocontacts 
hosting a single Co atom \cite{napoli:prl:13}.

Coordination complexes are of much interest in this regard. In particular, 
the family of planar organic molecules containing a transition metal (TM) 
center at its core are nicely suited for controlled experiments with scanning 
tunneling microscopy. The exchange-induced magnetic moment of the TM atom can 
be as high as $S=5/2$, but the strong coupling to the organic ligand usually 
quenches the spin into lower values \cite{williamson:jacs:92}. Crystal field 
theories can nicely explain the lowering of the high-spin state \cite{vanvleck:pr:32,*housecroft:book:04}. 
This phenomenon can also be properly described by standard implementations of density 
functional theory (DFT), i.e., by an effective one-electron approximation. Similarly, 
DFT can also explain charge transfer processes, typically between the TM and the surface, 
which can also quench the spin. However, often in these systems both one-body quenching 
and many-body screening processes coexist, making it very difficult to disentangle their 
respective contributions to the experimental signatures \cite{stepanow:prb:11}.

Here, by treating both screening and quenching on the same footing we elucidate the 
relevant mechanisms behind the experimental observations in a single Manganese Phthalocyanine 
(MnPc) absorbed on the (111) surface of Pb in the normal (i.e., not superconducting) 
phase \cite{fu:prl:07,franke:science:11}. This and similar systems have been recently 
studied both experimentally and theoretically 
\cite{fu:prl:07,franke:science:11,gao:prl:07,minamitani:prl:12,strozecka:prl:12,*Liu:SR:13}.
In contrast to previous theoretical work 
\cite{dasilva:prb:09,*korytar:jpcm:11,minamitani:prl:12,strozecka:prl:12,*Liu:SR:13} our approach
fully takes into account the electronic correlations and hybridization of the entire Mn $3d$-shell. 
This allows us to get the first complete picture of the Kondo effect and molecular quenching processes 
in a high-spin complex.
   
\begin{figure}
  \begin{minipage}[b]{0.4\linewidth}
    \includegraphics[width=0.8\linewidth]{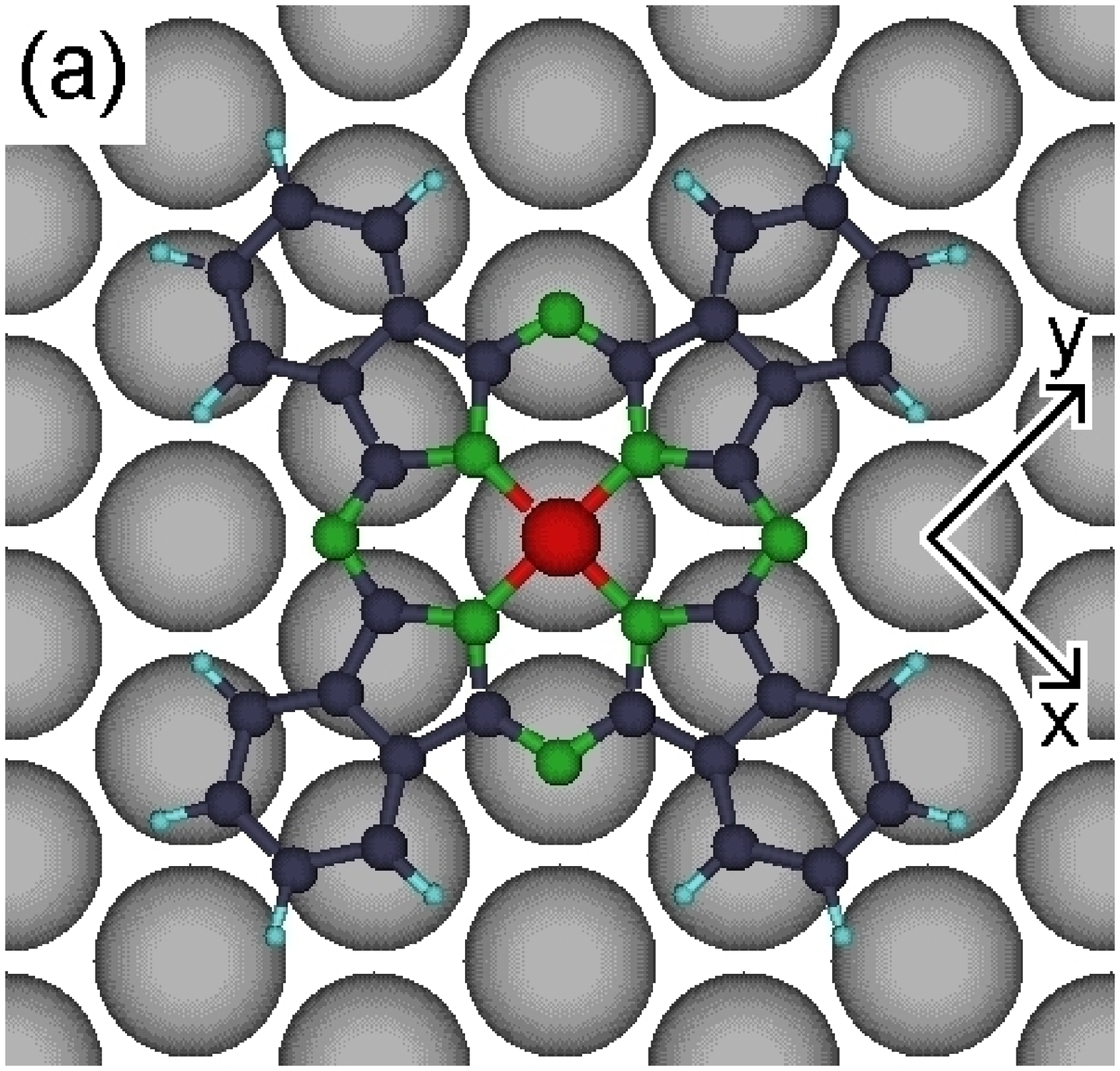} 
    \includegraphics[width=\linewidth]{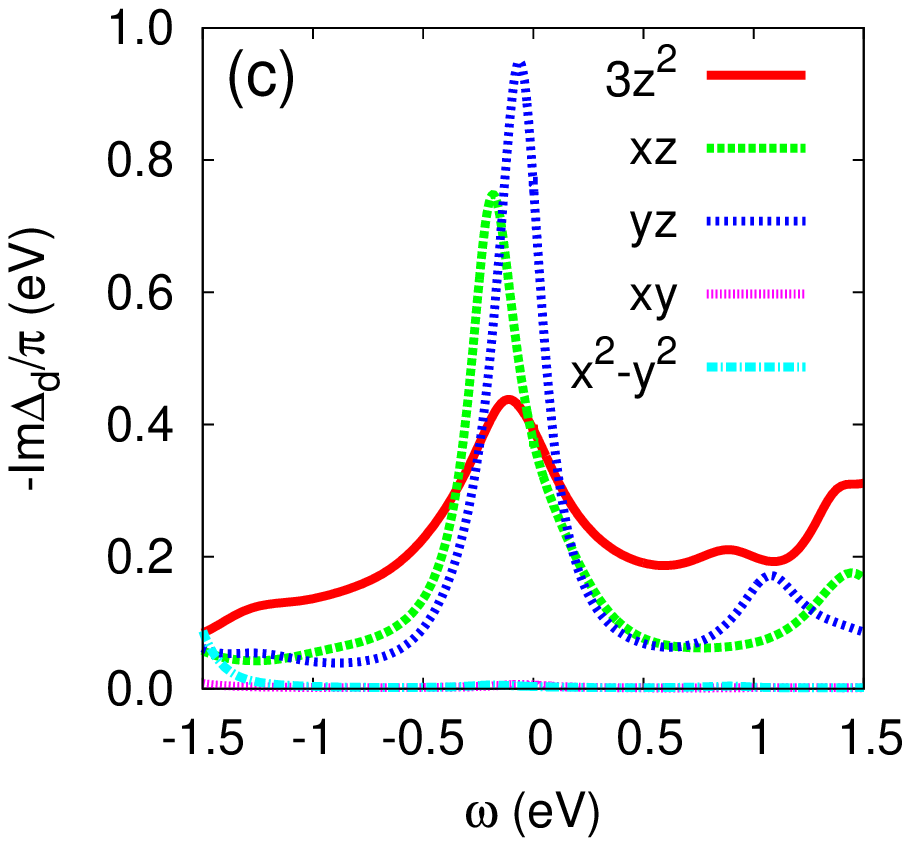} 
  \end{minipage}\hfill
  \begin{minipage}[b]{0.6\linewidth}
    \includegraphics[width=\linewidth]{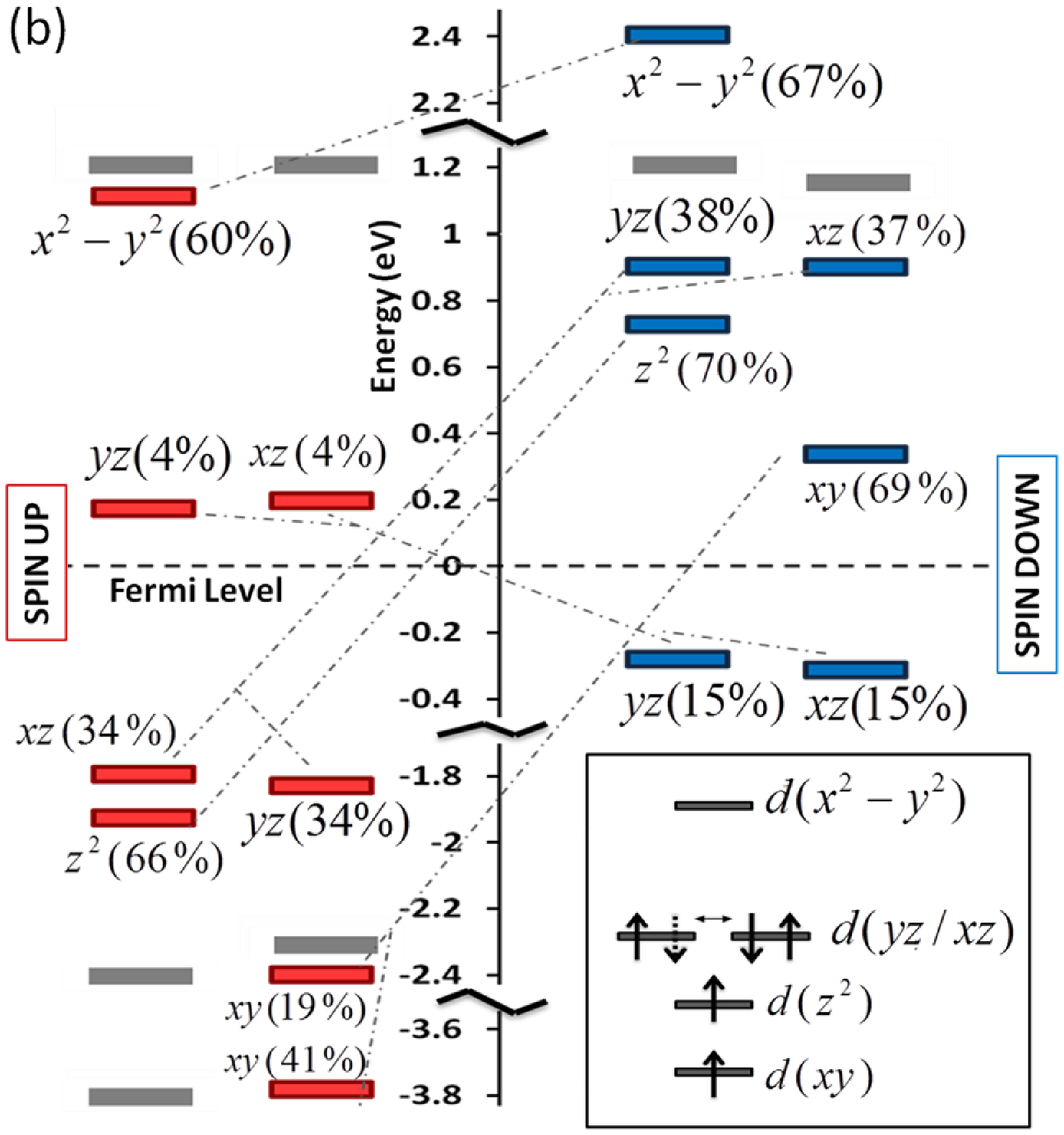} 
  \end{minipage}
  \caption{
    \label{fig:model}
    (a) Top view of a PcMn adsorbed on a Pb(111) surface in its more stable configuration.
    (b) Molecular orbital energy diagram as obtained from the projected spin-polarized KS Hamiltonian of the adsorbed molecule. 
    The Mn $3d$ orbital character (percentage) of each molecular orbital is shown. The inset shows the schematics of 
    the Mn $3d$ orbital energies and associated spins as considered in the OCA calculations.
    (c) Orbital resolved imaginary part of the hybridization function 
    for the Mn $3d$-orbitals as obtained from the GGA non-magnetic 
    electronic structure calculation. 
  }
\end{figure}

We consider a single MnPc adsorbed at the top site of a Pb(111) surface as shown in Fig. 1(a). 
We first relax the atomic structure, orientation, and distance of the molecule to the substrate which 
is represented by a cluster consisting of five atomic layers. This is done through the common Kohn-Sham 
(KS) approach to DFT using a standard GGA functional \cite{becke:pra:88}) as implemented in the Gaussian09 
package \cite{g09}. Next we embed the cluster consisting of the substrate and molecule (hereon called region C) 
into an effective semi-infinite bulk electrode model as implemented in the code ANT.G \cite{alacant,jacob:jcp:2011} 
which interfaces Gaussian09. The KS Green's function (GF) of the system C can now be obtained as 
$G^0_{\rm C}(\omega)=(\omega+\mu-H^0_{\rm C}-\Sigma_{\rm S}(\omega))^{-1}$ where $H^0_{\rm C}$ is the self-consistent 
KS Hamiltonian re-evaluated considering now $\Sigma_{\rm S}(\omega)$, which is the embedding self-energy describing 
the semi-infinite bulk electrode.

In order to capture many-body effects beyond the DFT level, we have applied the DFT+impurity 
solver method for nanoscopic conductors developed by one of us in earlier work \cite{jacob:prl:09,*jacob:prb:10a}. 
To this end the mean-field KS Hamiltonian is augmented by a Hubbard-like interaction term
$\hat{\mathcal{H}}_U=\sum_{\alpha\beta\gamma\delta\sigma\sigma^\prime} U_{\alpha\beta\gamma\delta}\,d_{\alpha\sigma}^\dagger d_{\beta\sigma^\prime}^\dagger d_{\delta\sigma^\prime} d_{\gamma\sigma}$
which accounts for the strongly interacting electrons of the Mn $3d$-shell. These are different 
from the bare interactions due to screening processes. The screened Coulomb 
interaction  within the Mn $3d$-shell, $U_{\alpha\beta\gamma\delta}$, has been determined using the constrained RPA approach
\cite{aryasetiawan:prb:04}.
We find that the matrix elements $U_{\alpha\beta\gamma\delta}$ are 
somewhat anisotropic with variations of up to 10\% between different orbitals.
For the {\it intra}-orbital Coulomb repulsion $U_{\alpha\alpha\alpha\alpha}$ we have
a mean value of 5.4~eV and for the {\it inter}-orbital Coulomb repulsion $U_{\alpha\beta\alpha\beta}\,(\alpha\ne\beta)$ 
a mean value of 4.1~eV. The orbital anisotropy of the direct repulsion
will be fully taken into account in our calculations.
The exchange matrix elements $U_{\alpha\beta\beta\alpha}$, which give rise to the
Hund's rule coupling $J_H$, also become somewhat orbital-dependent.
But here we simply set $J_H$ to the orbital-averaged exchange interaction  
$J_H\equiv\langle U_{\alpha\beta\beta\alpha}\rangle$ for which we find 0.65~eV.

The interacting Mn $3d$-shell coupled to the rest of the system (organic scaffold + surface) thus constitutes 
a so-called Anderson impurity model (AIM). The AIM is completely defined by the interaction matrix 
elements $U_{\alpha\beta\gamma\delta}$, the energy levels $\epsilon_d$ of the $3d$-orbitals and the 
so-called hybridization function $\Delta_d(\omega)$. The latter describes the (dynamic) coupling 
of the Mn $3d$-shell to the rest of the system and can be obtained from the KS GF \cite{jacob:prl:09}
as $\Delta_d(\omega)=\omega+\mu-\epsilon_d^0-[G^0_d(\omega)]^{-1}$ 
where $\mu$ is the chemical potential, $\epsilon_d^0$ are the KS energy levels 
of the $3d$-orbitals and $G^0_d(\omega)$ is the KS GF projected onto the 
$3d$-subspace. 
The energy levels $\epsilon_d$ are obtained from the KS levels, $\epsilon_d=\epsilon_d^0-E_{dc}$ 
where, as usual in DFT++ approaches \cite{lichtenstein:prb:98}, a double counting correction (DCC) 
has to be subtracted to compensate for the overcounting of interaction terms.
Here we employ the so-called fully localized or atomic limit DCC \cite{czyzyk:prb:94},
but generalized to the case of an anisotropic Coulomb repulsion:
$E_{dc}^\alpha=\sum_{\beta}U_{\alpha\beta\alpha\beta}\cdot\left(n_{\beta}-\frac{1}{2M}\right)-J_H\left(N_{3d}-1\right)/2$,
where $n_\alpha$ is the DFT occupation of orbital $\alpha$, $N_{3d}$ the total occupation of the Mn $3d$-shell,
and $M$, the number of correlated orbitals.

\begin{figure*}
  \includegraphics[width=\linewidth]{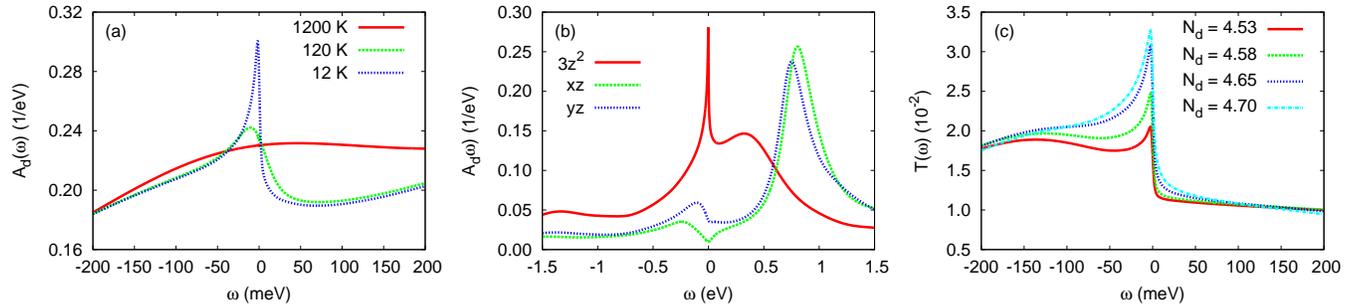}
  \caption{
    \label{fig:spectra}
    (a) Total spectral function of the Mn $3d$-shell near the Fermi level for different temperatures.
    (b) Orbital resolved spectral function of the Mn $3d$-shell at low temperature ($T\sim12$~K) in a larger energy window than in (a).
    (c) Transmission function for different filling $N_d$ of the four Mn $3d$-orbitals taken into account in the OCA calculation
    at low temperature ($T\sim12$~K).
  }
\end{figure*}

The AIM problem is now solved using the one-crossing approximation (OCA) \cite{haule:prb:01}.
This yields the electronic self-energy $\Sigma_d(\omega)$ which accounts for the electronic 
correlations of the $3d$-electrons due to strong electron-electron interactions. The 
{\it correlated} $3d$ GF is then given by $G_d=([G_d^0]^{-1}-\Sigma_d+E_{dc})^{-1}$. 
Correspondingly, the {\it correlated} GF for C is given by 
$G_{\rm C}=([G_{\rm C}^0]^{-1}-\Sigma_d+E_{dc})^{-1}$ where $\Sigma_d$ and $E_{dc}$
only act within the $3d$ subspace.
From $G_{\rm C}$ we can calculate the transmission function
$T(\omega)={\rm Tr}[\Gamma_{\rm T}\,G_{\rm C}^\dagger\,\Gamma_{\rm S}\,G_{\rm C}]$
where $\Gamma_{\alpha}\equiv i(\Sigma_{\alpha}-\Sigma_{\alpha}^\dagger)$ for
$\alpha = T,S$ describes the coupling of C to the STM tip (T) 
and to the semi-infinite Pb surface (S). Since the (small) 
voltage mainly drops between
tip and molecule, the transmission directly yields the differential conductance:
$\mathcal{G}(V)=(2e^2/h)T(eV)$.

The spin polarized KS spectrum of all molecular orbitals close to the Fermi energy is shown in Fig. 1(b) along with
their Mn $3d$-orbital character. The ones depicted in gray do not have  any Mn atomic character at all, being completely 
localized on the organic ligand. A strong  localization in the $x^{2}-y^{2}$, $xy$ and $z^{2}$ atomic orbitals is apparent, 
being signaled by a single molecular orbital (per spin) with strong atomic character. The $x^{2}-y^{2}$ appears as an empty 
molecular orbital, well above the Fermi level, the charge density in this orbital (see Table I) being only due to the 
contribution of many molecular orbitals with negligible participation of the atomic orbital.
The other two have a localized unpaired electron each. The third unpaired 
electron is shared between the $xz/yz$-orbitals. The localized character of this spin is masked due to the strong hybridization 
of these orbitals with the ligand, there being four molecular orbitals (per spin) with significant $xz/yz$ atomic character. 

In Fig. 1(c) we show the imaginary part of the hybridization function $\Delta(\omega)$
which describes the broadening of the Mn $3d$-levels due to the coupling to the substrate and to
the organic part of the molecule. We see that within an energy window of $\pm1.0$~eV around
the Fermi level, only three out of the five $3d$-levels are actually broadened ($3z^2,xz,yz$).
The $xy$- and $x^2-y^2$-orbitals, which are parallel to the surface, show no hybridization at 
all in this energy window. Outside this window (not shown), however, while 
the $xy$-orbital still does not show any significant coupling, the $x^2-y^2$-orbital presents 
a very large peak at -2.4~eV (and many small peaks) which indicates a strong coupling to the 
organic ligand. This is actually a manifestation of what crystal-field theory anticipates and 
the DFT calculation shows, as described in the previous paragraph.

Since the $x^2-y^2$-orbital is virtually empty and shifted to high energies, we exclude this orbital from 
the AIM model from now on.
The $3z^2$-, $xz$-, and $yz$-orbitals, being the only ones showing hybridization around the Fermi
level, are also the only ones susceptible to Kondo screening.  All three orbitals feature 
a strong peak in the hybridization function around the Fermi energy which stem from coupling 
to molecular orbitals in the organic ligand which, in turn, couple to the substrate. 
Note that due to symmetry reasons the direct coupling of the $xz$- and $yz$-orbitals to the substrate 
is strongly suppressed in the top position. In contrast, the $3z^2$-orbital also couples to the substrate 
directly, resulting in a flatter hybridization function. 

Let us now turn to the results of the OCA calculation for solving the generalized AIM problem. 
We find that, except for the $xy$-orbital, the orbitals of the Mn $3d$-shell are in a mixed-valence state. 
There are strong fluctuations between a five-fold degenerate atomic 
configuration with $N_d=4$ electrons and maximal spin $S_d=2$ where all orbitals are singly 
occupied and three four-fold degenerate atomic configurations with $N_d=5$ electrons and $S=3/2$
where one of the $3z^2$-, $xz$- and $yz$-orbitals is doubly occupied. 
This results in an average occupation of the four Mn $3d$-levels 
of $N_d\approx4.6$ electrons and an average total spin $\langle{S_d}\rangle\approx1.6$ close to $3/2$.
The extra half-electron stems from the emptied $x^2-y^2$-orbital and is shared among the $3z^2$-, $xz$-, 
or $yz$-orbitals.  This leads to strong charge fluctuations in these
orbitals (see Tab.~I) and thus quenching of their spin from $3/2$ to $\sim1$. 
The $xy$-orbital on the other hand is essentially singly occupied, thus carrying a spin-1/2. 
Note that the individual orbital channels are not in a mixed-valence situation as the individual
occupations are clearly below 1.5 and therefore the Kondo effect in individual orbitals is possible 
despite the Mn $3d$-shell as a whole being in a mixed-valence state \cite{nozieres:jphys:80}.
Although $\langle{S_d}\rangle$ is finite,
the expectation value for any of its projections is zero, since all states with $S_d^z=-S_d\ldots+S_d$
contribute equally. Also note that $\langle{S_d}\rangle\sim3/2$ {\it is not} the expectation value 
of the total spin of the system but only of the Mn $3d$-shell. The spin of the whole system is lower 
due to screening by the Kondo effect with the conduction electrons of the substrate and the organic rest of the molecule as we will see below.

\begin{table}
  \renewcommand{\arraystretch}{1.5}
  \begin{tabular*}{\linewidth}{l@{\extracolsep\fill}cccccc}
    {}  & $3z^2$ & $xz$ & $yz$ & $xy$   & $x^2-y^2$ & tot. \\
    \hline
    GGA     & 1.19 & 1.18 & 1.19 & 1.04 & 0.67   & 5.27 \\
    GGA+OCA & 1.32 & 1.13 & 1.16 & 1.04 & (0.67) & 5.32 
  \end{tabular*}
  \caption{Orbital occupations of Mn $3d$-shell as calculated with DFT on 
    the level of GGA and with OCA. Note that the $x^2-y^2$-orbital was
    not taken into account in the OCA calculation (see text). 
  }
\end{table}

Fig. 2a shows the spectral function of the Mn $3d$-shell for different temperatures. We see a 
sharp Kondo-peak developing right at the Fermi level when the temperature is lowered. As can be 
seen from Fig. 2b where we show the orbital-resolved spectral function on a larger energy scale 
than in Fig. 2a, the Kondo peak stems from the $3z^2$-orbital, the only orbital directly coupling 
to the substrate near the Fermi energy. 
Note that in the existing literature this orbital is
considered to be quenched \cite{fu:prl:07} and is excluded from correlated models \cite{bauer:prb:13}.
The Kondo temperature for this orbital is $T_K\sim100$~K. 
The $xz$- and $yz$-orbitals, on the other hand, each feature small bumps just below the Fermi level 
with a much larger width ($\sim 0.5$ eV) than the Kondo peak in the $3z^2$-orbital. 
We interpret these pronounced peaks in the hybridization function for the $xz$- and $yz$-orbitals 
as a result of these orbitals only coupling {\it via the organic ligands} to the substrate. In other words, 
these bumps suggests the formation of a many-body singlet state between the Mn $3d$-level and a molecular 
orbital in the organic rest of the molecule as in the zero-bandwidth Anderson impurity model 
(see e.g. App. of ref.~\onlinecite{hewson:book:97}). In this model the formation of the total spin-singlet
state between the strongly interacting impurity level and a {\it single} non-interacting bath level
gives rise to two strongly renormalized resonances below and above the Fermi level. These resonances 
are precursors of the Kondo peak which develops as more and more bath levels are added to the model. 
Therefore we can think of the spin in the $xz$- and $yz$-orbitals as being screened due to 
the formation of a many-body singlet state by strong coupling with the organic ligand. 
The spin-$1/2$ in the $xy$-orbital, on the other hand, remains unscreened due to lack of
 hybridization with the substrate or molecule near the Fermi level. 
This is in contrast to existing claims where this orbital is considered to be screened and responsible 
for high-energy Kondo features \cite{fu:prl:07}.

Therefore we are dealing here with a $S=3/2$ {\it underscreened} Kondo effect where only the spin $S\approx1$
within the $3z^2$-, $xz$- and $yz$-orbitals is screened, leaving a residual spin-1/2 in the Mn $xy$-orbital 
which may lead to so-called singular Fermi-liquid behavior \cite{coleman:prb:03}.
Only the screening of the spin within the $3z^2$-orbital 
gives rise to a Kondo resonance while no significant low-bias experimental signatures are expected from the strongly 
coupled spin in the $xz/yz$-orbitals. 
Since the Kondo temperature of the $3z^2$ channel is too high, it is not unrealistic to attribute the 
Shiba peaks in Ref. \cite{franke:science:11} to a low-energy scale Kondo screening of the $xy$-orbital 
in a lower symmetry experimental situation \cite{bauer:prb:13}.

The Kondo resonance appearing in the spectral function of the Mn $3d$-shell for low temperatures is 
somewhat asymmetric. This is mainly a result of two effects: On the one hand charge fluctuations
in the $3z^2$-orbital make the Kondo peak asymmetric due to the proximity of the upper Hubbard peak.
On the other hand the modulation of the hybridization function due to the coupling to the organic ligands
near the Fermi level further enhances this asymmetry. The bumps in the spectral function of the $xz$- and 
$yz$-orbitals on the other hand do not have a significant contribution to the asymmetry of the Kondo peak 
due to their small spectral weight. This asymmetry of the Kondo peak in the spectral function is even more 
enhanced in the tunneling spectra as can be seen in Fig.~2c where we show the tunnel transmission $T(\omega)$ 
calculated for a Pb tip positioned above the Mn atom in a distance of 5~\r{A}. The reason for 
this further enhancement is the modulation of the Mn $3d$ spectral function by the DOS of the Pb tip and the 
Pb substrate. For a Au tip we actually find that the tunnel spectra (not shown) are a little bit less asymmetric. 
Therefore the peak in the tunnel spectra just stems from the Kondo peak in the $3z^2$-orbital. In fact, a sharp
Kondo peak in either of the $xz$- or $yz$-channel would rather give rise to a dip in the tunneling, but not 
to a peak since the direct tunnel matrix elements between the tip and these two orbitals vanish for symmetry reasons.

Charge fluctuations usually have a strong effect on the Kondo screening. In Fig. 2(c) we show the
effect of altering the occupation of the Mn $3d$-shell on the tunneling spectra 
by shifting the Mn $3d$-levels by a few decimal eV. We see that the shape of the Kondo resonance 
and in particular its width is strongly affected by the slight changes in the occupation of the 
Mn $3d$-shell. In fact our calculated lineshapes reproduce very well the variation of lineshapes 
measured in recent experiments. Hence we conclude that the experimentally observed variation in 
lineshapes for different PcMn molecules on the Pb(111) surface \cite{fu:prl:07,franke:science:11}
is likely due to slight changes in the occupation of the Mn $3d$-shell induced by slight variations 
in the structure or environment of the molecule in the experiments. 

In summary, we have studied the correlated electronic structure of a MnPc adsorbed on the Pb(111) surface, 
fully taking into account the strong electronic correlations originating from the Mn $3d$-shell. Our results 
show that the adsorption does not essentially modify the total spin $S=3/2$ of the molecule, which is distributed 
among four of the five $3d$-orbitals. This finding is in stark contrast to previous works which assume/find 
a spin-$1$ \cite{bauer:prb:13} or even a spin-$1/2$ state due to strong quenching with the substrate and organic ligand 
\cite{fu:prl:07}. We further find that the experimentally observed asymmetric Kondo resonance in this system 
\cite{franke:science:11,fu:prl:07} is due to an underscreened Kondo effect where a spin-$1/2$ in the Mn $3d$-shell 
remains unscreened. The Kondo resonance in the tunnel spectra actually stems from only one of the Kondo-screened 
orbitals. Its peculiar lineshape arises from the modulation of the hybridization function due to strong coupling 
to the organic ligand, not being necessary to invoke the superposition of two Kondo peaks with different 
Kondo temperatures as done in Ref. \cite{franke:science:11}. 

This work was supported by MICINN under Grants
No. FIS2010-21883 and No. CONSOLIDER CSD2007-
0010. M. S. acknowledges computational support from the
CCC of the Universidad Aut\'onoma de Madrid.
We are also grateful to K. Haule for providing 
us with the OCA impurity solver.

\bibliographystyle{apsrev4-1}
\bibliography{kondo}

\end{document}